\documentstyle[aas2pp4,12pt]{article}

\def\mm{M\'endez }
\def\cygx1{Cyg\,X--1}
\def\gx339{GX\,339--4}
\def\1705{4U\,1705--44}
\def\x0614{4U\,0614+09}

\begin{document}

\title{Measurement of Hard Lags and Coherences in the
X-Ray Flux of Accreting Neutron Stars and Comparison with Accreting
Black Holes}

\author{Eric C. Ford\altaffilmark{1}, Michiel van der
Klis\altaffilmark{1}, Mariano \mm\altaffilmark{1,2}, Jan van
Paradijs\altaffilmark{1,3}, Philip Kaaret\altaffilmark{4}}

\authoremail{ecford@astro.uva.nl}
\altaffiltext{1}{Astronomical Institute, ``Anton Pannekoek'',
University of Amsterdam, Kruislaan 403, 1098 SJ Amsterdam,
The Netherlands}
\altaffiltext{2}{Facultad de Ciencias Astron\'omicas y Geof\'{\i}sicas,
Universidad Nacional de La Plata, Paseo del Bosque S/N, 1900 La Plata,
Argentina.}
\altaffiltext{3}{University of Alabama in Huntsville, Department
of Physics, Huntsville, AL 35899}
\altaffiltext{4}{Harvard--Smithsonian Center for Astrophysics, 
60 Garden Street, Cambridge, MA 02138}

~~

\centerline{\bf {TO APPEAR IN ApJ LETTERS}}

\begin{abstract}

Using the Rossi X-ray Timing Explorer we have measured lags of the 9
to 33 keV photons relative to the 2 to 9 keV photons in the timing
noise between 0.01 and 100 Hz in the accreting neutron stars \x0614
and \1705.  We performed similar measurements on the accreting black
hole candidates \cygx1 and \gx339 as a comparison.  During the
observations these sources were all in low (hard) states.  We find
phase lags of between 0.03 and 0.2 radians in all these sources, with
a variation in frequency much less than expected for a lag constant in
time. We also measure a coherence consistent with unity in all
sources.  As already noted for the black hole candidates, these data
are inconsistent with simple Comptonization models invoking a constant
time delay. Comptonization in a non-uniform medium can perhaps explain
the lags. However, the magnitudes of the lags imply that the hot
electron gas extends to more than $10^3$ Schwarzschild radii. This may
constitute an energy problem.  We argue that while a large hot cloud
is possible for black holes which may hide some of their accretion
energy in advection, such a distribution may not be possible neutron
stars where all the accretion energy is eventually released at the
neutron star surface. This casts doubt on the Comptonization model,
though the energy problem may be resolved for example by a wind from
the inner disk.

\end{abstract}


\section{Introduction}

The X--ray properties of black hole X--ray binaries in the ``low
state'' are strikingly similar to those of atoll sources (accreting
low magnetic field neutron stars) in the ``island state'' (van der
Klis 1994a,b). Both emit X--ray spectra that are dominated by a hard
power law component, $N(E) \propto E^{-\alpha}$, whose photon index
$\alpha$ increases (i.e. whose spectra soften) as the mass accretion
rate, $\dot M$, increases (Tanaka \& Lewin 1995; Barret \& Vedrenne
1994, van Paradijs \& van der Klis 1994). The power density spectra of
the X--ray flux variations of both types of sources are dominated by a
broad-band component, which is flat below a break frequency, $\nu_{\rm
b}$, and turns to a power law, $P(\nu) \propto \nu^{-\beta}$ above
$\nu_{\rm b}$, with $\beta \sim 1$ just above the break (see van der
Klis 1995).

As $\dot M$ increases, so does $\nu_{\rm b}$; the observed values of
$\nu_{\rm b}$ are similar for the black hole and neutron star systems
and cover the range 0.01 to 30 Hz. As $\nu_{\rm b}$ varies, the
high--frequency part of the power density spectrum tends to remain the
same (Belloni \& Hasinger 1990; Miyamoto et al. 1992), although this
relation is not always exact in detail. Generally it is observed
that the power level of the flat part of the power density spectrum is
anti-correlated with $\nu_{\rm b}$ (and $\dot M$) in black hole
candidates (\mm \& van der Klis 1997 and references therein) and in
neutron stars (Yoshida et al. 1983, Prins \& van der Klis 1997, Ford
\& van der Klis 1998, \mm et al. 1998).

Another characteristic of the low states of black hole X--ray binaries
(BHXB) is the phase lag of high-energy photons with respect to
low-energy photons (``hard lags'') in flux variations covering time
scales between $\sim 10^{-2}$ to $\sim 10^2$ s (see Miyamoto et
al. 1992). This phase lag is roughly constant over this range, at a
level of order 0.1 radians, and does not vary much with $\nu_{\rm b}$
(i.e. $\dot M$) for a given source or between sources.  These phase
lags have been interpreted (eg. Payne 1980) as travel time differences
of photons in a hot Compton scattering electron gas (e.g. Sunyaev \&
Tr\"{u}mper 1979), which is also the medium in which the power law
photon spectrum is thought to be formed (e.g. Hua \& Titarchuk 1995;
Dove et al. 1997). In accreting neutron stars, lags have been measured
in the kilohertz oscillation signals (Vaughan et al. 1997;
1998). These very small lags put stringent constraints on the physical
size of any cloud which might Comptonize the photons in the fast
signals.

It is clearly of interest to know whether the similarities between
island--state atoll sources and low--state BHXBs extends to their time
lag properties.  We present here the results of a time lag analysis of
two atoll sources at frequencies below 100 Hz and compare the results
to an identical analysis of two BHXBs.

In Section 2 we describe the observations, and data analysis. The results 
of our analysis are presented in Section 3. They are discussed in the 
framework of the Comptonization model in Section 4. We conclude that 
this model has fundamental problems.

\section{Observations and Analysis}

We have used data from the Rossi X--ray Timing Explorer (RXTE)
proportional counter array (PCA) obtained during the first three
observing cycles (see Table~\ref{tbl:obs}). We choose four sources:
\cygx1 and \gx339, which are thought to contain black holes (Tanaka \&
Lewin 1995), and \x0614 and \1705, which are atoll sources showing
X--ray bursts and therefore contain neutron stars. In these
observations the BHXBs are in the `low' state and the atoll sources
are in the `island' state. They have similar Fourier power spectra
which can be fitted by a broken power law, with break frequencies of
0.18, 0.07, 1.8, and 3.6 Hz for \cygx1, \gx339, \1705, and \x0614
respectively, and rms fractions of 29\%, 36\%, 22\%, and 31\% (2 to 9
keV, 0.01 to 100 Hz). The mean count rates are 4973, 586, 580, and 228
counts s$^{-1}$ (2 to 9 keV, background subtracted for the whole PCA),
with small long term variability.

To quantify the difference in variability between two energy bands we
use the cross spectrum defined as $C(j) = X_1(j)^{*}X_2(j)$, where $X$
are the measured complex Fourier coefficients for the two energy bands
at a frequency $\nu_j$ (van der Klis et al. 1987; for information on
cross correlation analysis see Vaughan et al. 1994; Vaughan \& Nowak
1997; Nowak et al. 1998). The phase lag, $\phi$, between the
signals in the two bands is given by the argument of $C$ (its position
angle in the complex plane). The corresponding time lag at Fourier
frequency, $\nu$, is simply $\phi/2\pi\nu$.  We calculate an
average cross--vector, $C$, by averaging over multiple spectra and
binning in frequency, and then find $\phi$.  We calculate the
error in $\phi$ from the observed variance of the $C$ values in
the real and imaginary directions. The resulting error is similar to,
but slightly larger than, that derived from counting statistics
(Vaughan et al. 1994; Cui et al. 1997), or (equivalently) from the
coherence function uncorrected for counting statistics (Nowak et
al. 1998).

We calculate cross--spectra from data intervals of 256 sec and use a
Nyquist frequency of 2048 Hz, which is high enough to avoid binning
effects that dominate at frequencies above half the Nyquist frequency
(Crary et al. 1998). As the resulting $\phi$ depends somewhat
on the choice of energy band, becoming larger for wider energy
separations, we choose the energy bands to be closely similar.  We
choose two energy bands defined as effectively 2 to 9 keV and 9 to 33
keV. These bands correspond to PCA channels 0--35 and 36--127 for the
\cygx1 observations, and channels 0--25 and 26--87 for the other data
which were taken after a detector gain change in March 1996.  We
correct for deadtime effects by subtracting from $C$ an average value
obtained from 800 to 1024 Hz where Poisson noise dominates (van der
Klis et al. 1987). This correction is close to negligible.  A positive
value of $\phi$ corresponds to a lag of the hard band relative
to the soft band, a result we have confirmed by analyzing test signals
and looking at data from the accreting millisecond X--ray pulsar
SAX\,J1808.4--3658.

\section{Results}

\begin{figure*}
\figurenum{1}
% need rescaling in preprint mode
\epsscale{2.0}
\plotone{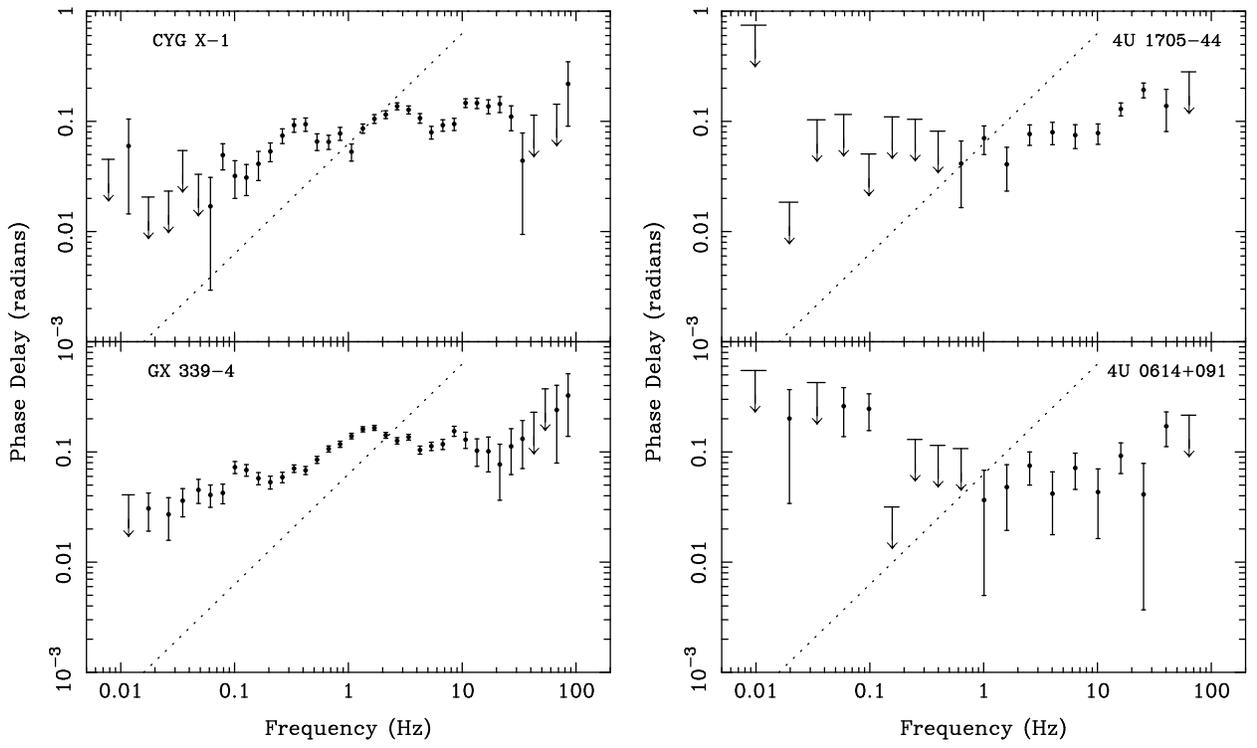} 
\caption{Phase delays between signals in the 2--9 keV and 9--33 keV
bands for two black hole candidate binaries in the low state (left)
and two neutron star binaries in the `island state' (right). The
dotted lines show constant time delays of 10 msec.  Measurements which
are less than 1$\sigma$ significant are plotted as 95\% confidence
upper limits.}
\label{fig:pdel}
\end{figure*}

The results of the phase delay analysis are summarized in
Figure~\ref{fig:pdel}. In all four sources the hard photons lag the
soft photons significantly.  Even in \x0614, where the statistics are
the worst, the lag is detectable with a significance greater than
5$\sigma$ in the 1 to 10 Hz range. Averaging over the 0.01 to 100 Hz
range, the $\phi$ values are $0.077\pm0.003$, $0.093\pm0.003$,
$0.092\pm0.011$, $0.067\pm0.016$ radians for \cygx1, \gx339, \1705,
and \x0614, respectively. In \1705 and \x0614, the lags are
significant only above 1 Hz; over the range 0.01--1 Hz $\phi$
is $0.030\pm0.014$ and $0.047\pm0.021$ radians respectively.

The significance of the phase lags is a factor of 3 to 10 better for
the BHXBs than for the atoll sources.  In \cygx1 this is because of
the high count rate. In \gx339 it is a result of the high rms fraction
of the noise and long integration time.  We calculate the expected
significances which are a function of the rate, rms fraction,
observing time and coherence (Nowak et al. 1998).  Compared to \x0614,
the significances should be better by factors of 13, 14, and 3 in
\cygx1, \gx339, and \1705, respectively, similar to what we have
observed.  There is structure visible for $\phi$ as a function of
frequency in the BHXBs (Miyamoto et al. 1992), in particular $\phi$
increases somewhat with frequency. The lags are not, however,
consistent with a constant time lag as shown by the dotted lines in
Figure~\ref{fig:pdel}.

\begin{figure*}
\figurenum{2}
% need rescaling in preprint mode
\epsscale{2.0}
\plotone{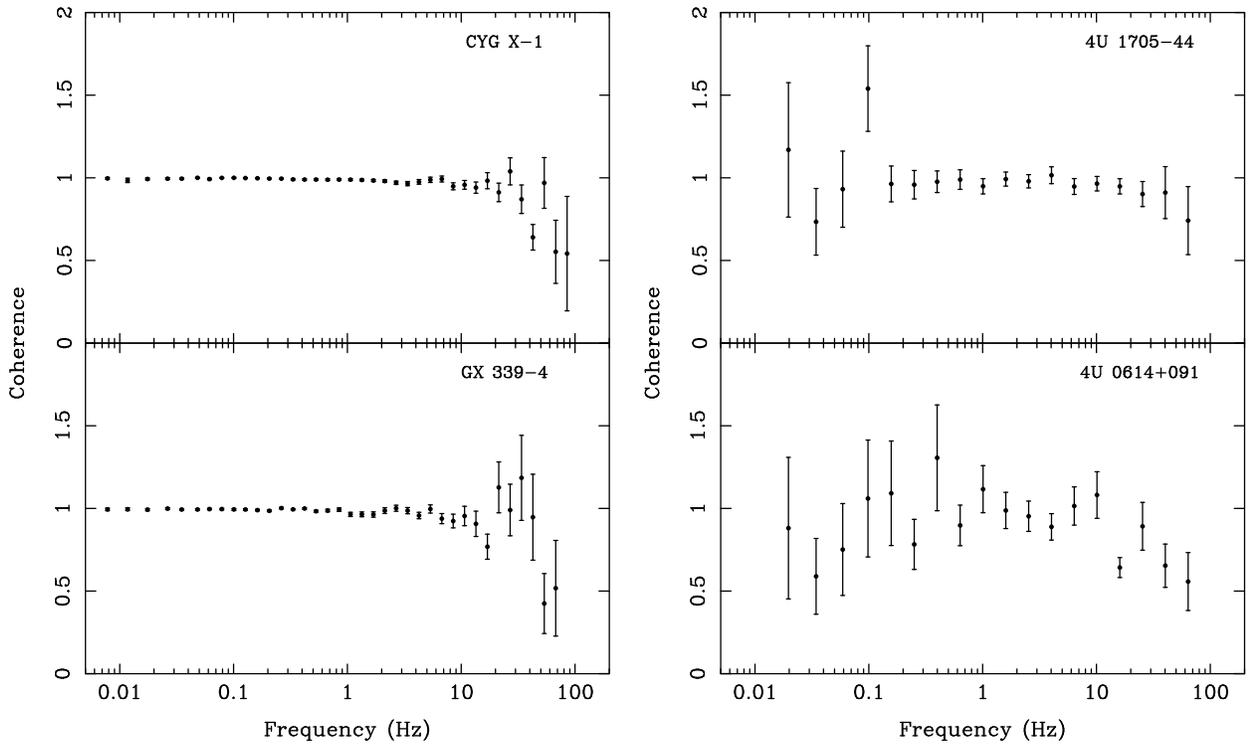} 
\caption{Poisson noise corrected coherence functions between the energy
bands 2--9 keV and 9--33 keV.}
\label{fig:g2}
\end{figure*}

In addition to the phase delay, we measure the coherence, defined by
$\gamma^2=|<C(j)>|^2/(<|X_1(j)|^2><|X_2(j)|^2>)$ (Vaughan \& Nowak
1997). The angle brackets represent averages over multiple
cross--spectra and Fourier frequencies. If the cross vectors have the
same phase angles, $\gamma^2$ is unity; if the phases are random,
$\gamma^2$ is zero. We corrected $\gamma^2$ for the contribution from
Poisson noise (see Vaughan \& Nowak 1997); this correction
significantly changes the calculated values. We calculate the error of
the noise--corrected $\gamma^2$ according to the prescription of
Vaughan \& Nowak (1997) in the high--coherence, high--variability regime
applicable to our data below 100 Hz.  Figure~\ref{fig:g2} shows the
results. The coherence is consistent with unity for all sources.

\section{Discussion}

We have presented the first measurements of phase lags and coherence
in the band limited noise below 100 Hz in accreting neutron stars. We
find that the phase lags of the atoll sources \1705 and \x0614 in the
island state are very similar to those of BHXBs in the low state as
reported here and by previous authors (Page et al. 1981; Miyamoto et
al. 1988; Miyamoto et al. 1992; Nowak et al. 1998; Pottschmidt et
al. 1998). In the noise at frequencies between 0.01 and 100 Hz, hard
X--rays lag soft X--rays. The average phase delays that we measure in
this frequency range are 0.09 and 0.07 radians for \1705 and \x0614,
and 0.08 and 0.09 radians for the low-state BHXBs \cygx1 and
\gx339. The lags are not consistent with a constant time delay.

We discuss our results within the framework of the idea that the
X--ray photons in the power-law spectral component are the result of
Compton upscattering of low-energy seed photons, probably produced in
the accretion disk, by a very hot electron gas.  This implies that the
energy emitted in X--rays first resides in energetic electrons, which
subsequently lose part of their energy to low-energy photons in the
scattering processes. In this Comptonization model, a low energy photon
on average gains energy in each scattering event.  Higher energy
photons, on average, are the result of more scatterings.  Therefore
the higher energy photons emerge later than the lower energy photons
produced simultaneously, since the total path length they traverse is
larger due to more scatterings.  The measured unity coherence in all
sources argues for such a model as opposed to scenarios where signals
at different energies are produced in disconnected regions (Vaughan \&
Nowak 1997).

In simple versions of Compton scattering in a uniform cloud, one
expects a constant time lag between photons in different energy bands
independent of frequency (Wijers et al. 1987), i.e.  the phase lag,
$\phi$, is expected to increase proportionally to the Fourier
frequency, $\nu$, at least when the Fourier period is larger than the
typical time delay.  This is inconsistent with the data as noted
previously.  The recent work of Kazanas, Hua \& Titarchuk (1997;
KHT97) and Hua, Kazanas \& Titarchuk (1997; HKT97), however, shows
that a scattering medium with a radial density gradient can produce
phase lags roughly constant in frequency.

We also note that in the atoll sources there are also quasi-periodic
oscillation signals in excess of 1000 Hz which are thought to originate
very close to the neutron star. The upper limits to any hard phase
lags in these signals is small (Vaughan et al. 1998). For a medium
which induces constant time delays, the measured lags of these fast
signals should have been large.

The size of the Comptonizing region is large. Our measurements of a
0.02 sec lag at 1 Hz for the neutron stars and 0.2 sec at 0.1 Hz for
the black holes correspond to a size of order $10^4$ km for moderate
optical depths.  Even in the KHT97 and HKT97 models with non--constant
density, the scattering medium extends up to at least $10^4 \rm R_{\rm
S}$, where $\rm R_{\rm S}$ is the Schwarzschild radius.  At the same
time a large fraction of the total luminosity is released in hard
X--rays which are apparently produced by the Comptonizing medium.
Relative to the flux in the 0.1--100 keV band, in our observations
\cygx1 emits 65\% of its energy above 9 keV; in \1705 this fraction is
25\%.  The question then arises: how can a substantial fraction of the
emitted X--ray luminosity, which must originate from the conversion of
gravitational potential energy into heat close to the compact object,
reside in hot electron gas at distances of order $10^3 \rm R_{\rm S}$
away from the compact object? Energy production around accreting
neutron stars was considered early on (Zel'dovich \& Shakura 1969).
The energy problem discussed above has been raised by recent authors
(e.g. Stollman et al. 1987).

For the black hole systems one might argue that the energy budget in
hot electron gas at distances of order $10^{9}$ cm {\em is} sufficient
to account for a substantial part of the observed X--ray luminosity
via Compton upscattering. The required mass accretion rate then has to
be of order $10^{-7}$ M$_{\odot}$, i.e., several orders of magnitude
above the value inferred from the observed X--ray luminosities of
$\sim 10^{37}$ erg/s. The consequence of this is that the accretion
flow within $\sim 10^{9}$ cm of the black hole must have a very low
radiative efficiency, which suggests that it is advection-dominated
(see, e.g., Narayan \& Yi 1994).  In this picture it would be the
advection dominated accretion flow (ADAF) itself which provides the
site for Comptonization. (Note that at $10^{9}$ cm the virial
temperature is already $\sim 200$ keV, in principle sufficient for the
formation of a hard power law).

However, a similar ADAF flow picture cannot apply to the atoll
sources.  These sources cannot have such a high accretion rate, since
they have no way of hiding the corresponding photon luminosity
liberated near the neutron star surface.  The energetics is a problem
for disk configurations where the losses from Comptonization are
expected to exceed the energy locally available from gravitational
release or radiative heating (Shibazaki et al. 1988). KHT97 suggest
that the Comptonizing region is a quasi-spherical flow preheated by
radiation from a central region (references in KHT97, Zel'dovich \&
Shakura 1969), perhaps a hot boundary region (Titarchuk, Lapidus \&
Muslimov 1998). Transporting a large fraction of energy to large radii
may also be accomplished by a wind blown from the inner disk
(Achterberg 1998).  Magnetic fields offer another possibility for
transporting energy from near the compact star to larger radial
distances (Stone et al. 1996).  To maintain the Comptonizing region in
the neutron star systems a large fraction (at least 25\%) of the total
accretion power must be efficiently transported from its point of
release, within a few tens of km of the neutron star, out to distances
of order $10^9$ cm.

To summarize, we have shown that the phase delays and coherences
measured in neutron stars are the same in X-ray binaries which likely
contain black holes.  If these lags are from Comptonization, the size
of the scattering medium is large which may pose a problem for the
energetics.  Delays generated by processes other than Comptonization
may have to be considered as has been done recently (Bottcher \& Liang
1998). The correct model must describe both accretion onto black holes
and neutron stars.

% Acknowledgments:
We thank Stefan Dieters and Jeroen Homan for comments on the
manuscript. ECF acknowledges support by the Netherlands Foundation
for Research in Astronomy with financial aid from the Netherlands
Organization for Scientific Research (NWO) under contract numbers
782-376-011 and 781-76-017.  M. M. is a fellow of the Consejo Nacional
de Investigaciones Cient\'{\i}ficas y T\'ecnicas de la Rep\'ublica
Argentina. JvP acknowledges support from NASA grants NAG5-4482,
NAG5-7382 and NAG5-7415. PK acknowledges support from NASA grants
NAG5-7405, NAG5-7407, and NAG5-7477.

%%%%%%%%  TABLE 1: Observations  %%%%%%%%%%%

\begin{deluxetable}{lcccc}
\tablenum{1}
\tablewidth{40pc}
\tablecaption{RXTE Observations}
 
\tablehead{ 
\colhead{Source} & \colhead{Start Time} & \colhead{Duration} &
\colhead{$T_{\rm res}$} & \colhead{${\rm N}_{Chan}$} \nl
\colhead{} & \colhead{(UTC)} & \colhead{(ksec)} &
\colhead{($\mu$sec)} & \colhead{} \nl
}

\startdata
Cyg X--1    & 12 Feb 1996 12:45 &  3.1 & 16  &  16 \nl
GX 339--4   &  3 Feb 1997 15:56 & 13.1 & 125 & 128 \nl
            & 10 Feb 1997 15:40 & 11.8 & 125 & 128 \nl
            & 17 Feb 1997 18:17 &  7.9 & 125 & 128 \nl
4U 1705--44 &  1 Apr 1997 13:26 & 12.0 & 125 &  64 \nl
4U 0614+09  & 22 Apr 1996 19:19 & 26.9 & 125 &  64 \nl
            & 28 Aug 1998 19:30 & 13.6 & 125 &  64 \nl

\tablecomments{Listed are the start time (in Universal Time,
Coordinated) of each observation and the duration of data used. We use
`Event mode' data from the PCA with time resolution $T_{\rm res}$,
and a number of channels ${\rm N}_{Chan}$.}

\enddata
\label{tbl:obs}
\end{deluxetable}

%%%%%%%%  FIGURES:

\end{document}